\documentclass[journal=jacsat,manuscript=article]{achemso}
\usepackage{chemformula} 
\usepackage{sectsty}
\usepackage{graphicx} 
\usepackage{lastpage}
\usepackage[format=plain,justification=justified,singlelinecheck=false,font={stretch=1.125,small,sf},labelfont=bf,labelsep=space]{caption}
\usepackage{float}
\usepackage{fancyhdr}
\usepackage{fnpos}
\usepackage[english]{babel}
\usepackage{droidsans}
\usepackage{charter}
\usepackage[T1]{fontenc}
\usepackage[utf8]{inputenc}
\usepackage{textcomp}
\usepackage{kotex}
\usepackage{amsfonts}
\usepackage{amsmath,bm,amssymb}

\author{Do Hoon Kiem}
\author{Min Yong Jeong}
\author{Hongkee Yoon}
\author{Myung Joon Han}
\affiliation[KAIST]
{Department of Physics, Korea Advanced Institute of Science and Technology (KAIST), Daejeon 34141, Korea}
\email{mj.han@kaist.ac.kr}

\title[An \textsf{achemso} demo]
  {Strain engineering  and the hidden role of magnetism in monolayer VTe$_2$}


\begin{document}
 
\begin{abstract}
		Two-dimensional transition-metal dichalcogenides have attracted great attention recently. Motivated by a recent study of crystalline bulk VTe$_2$, we theoretically investigated the spin-charge-lattice interplay in monolayer VTe$_2$. To understand the controversial experimental reports on several different charge density wave ground states, we paid special attention to the ‘hidden’ role of antiferromagnetism as its direct experimental detection may be challenging. Our first-principles calculations show that the 4$\times$1 charge density wave and the corresponding lattice deformation are accompanied by the ‘double-stripe’ antiferromagnetic spin order in its ground state. This phase has not only the lowest total energy but also the dynamical phonon stability, which supports a group of previous experiments. Interestingly enough, this ground state is stabilized only by assuming the underlying spin order. By noticing this intriguing and previously unknown interplay between magnetism and other degrees of freedom, we further suggest a possible strain engineering. By applying tensile strain, monolayer VTe$_2$ exhibits phase transition first to a different charge density wave phase and then eventually to a ferromagnetically ordered one.
\end{abstract}

\section{Introduction}

Two-dimensional (2D) materials and their magnetism have recently attracted great attention \cite{burch_magnetism_2018, huang_layer-dependent_2017,gong_discovery_2017, gibertini_magnetic_2019, mak2019probing}. In particular, transition-metal dichalcogenides display many fascinating properties including topological band structure \cite{qian_quantum_2014, keum2015bandgap}, intriguing optical and `valley-tronic' character \cite{xu_spin_2014,sie_valley-selective_2015, zhong2017van,wang_electronics_2012,wang_colloquium_2018} as well as magnetism \cite{zhou2012tensile, bonilla2018strong, guo2017modulation} and superconductivity \cite{sipos2008mott, ugeda2016characterization}. A characteristic feature of this family of materials is that multiple phases compete or collaborate with one another, and thereby exhibit many interesting phenomena. The strong inter-connection of spin, charge, and lattice degree of freedom is behind those observations, which poses outstanding physics problems and simultaneously provides great potential for future applications.

Recently, a notable case of polymorphism has been highlighted in VTe$_2$ \cite{won2020polymorphic}. From the high-quality single crystalline sample of bulk VTe$_2$, Won and coworkers showed that two different types of charge density waves (CDWs) can be stabilized in close collaborations with two different spin ordering patterns \cite{won2020polymorphic}. In particular, the energetically favorable lower-defect phase of so-called `type 2' CDW becomes stable only when it comes with a certain antiferromagnetic (AFM) order although the direct experimental identification could not be conducted. Given that the experimental verification of AFM order is challenging especially for many of nano-sized samples, the `unseen' or `hidden' role of magnetism can be of critical importance to determine the ground state configuration of ultra-thin transition-metal dichalcogenides.

Here we note that such an intriguing role of magnetism can also be crucial in VTe$_2$ monolayer for which the controversial results have been reported. Experimentally, several different types of CDW were observed. Wang et al. reported $4\times4$ CDW from their angle-resolved photo-emission spectroscopy (ARPES) and low-energy electron diffraction (LEED) experiment \cite{wang_evidence_2019}, which is consistent with independently-performed scanning tunneling microscope (STM)  studies \cite{liu_multimorphism_2020, miao_real-space_2020, wu_orbital-collaborative_2020, coelho_monolayer_2019,wong_metallic_2019}. However, other STM experiments observed the different CDW phases such as $4\times1$ and $2\sqrt{3}\times2\sqrt{3}$  \cite{wu_orbital-collaborative_2020,liu_multimorphism_2020}. Liu et al. also reported  $5\times1$ CDW at the grain boundary \cite{liu_multimorphism_2020}. In addition, for bilayer and a-few-layer-thick films, $2\times1$ and $4\times1$ patterns were reported \cite{ma_evidence_2012,dai_multiple_2019, coelho_monolayer_2019}. Theoretical investigations were not quite useful to resolve this controversial issue. Previous density functional theory (DFT) calculations reported that the most stable CDW configuration for monolayer VTe$_2$ is the bulk-like $3\times1$ which is not consistent with any of previous experiments \cite{coelho_monolayer_2019, wong_metallic_2019}. Thus the situation is still quite controversial even after the intensive theoretical and experimental studies. It demonstrates the difficulty of characterizing this type of materials in which the multiple degrees of freedom can coexist and compete particularly in the nano-meter scale.

Motivated by the fact that magnetism has largely been ignored in this material VTe$_2$, we performed the extensive DFT calculations with serious attention to the possible underlying magnetism. We investigated all of the reported CDW phases together with the possible spin orders that have been studied in bulk VTe$_2$ \cite{won2020polymorphic}. We found that, only when the underlying spin order is taken into account, the stable CDW phase can be obtained in terms of energetics and dynamic stability. Our calculations show that the ground state CDW of VTe$_2$ monolayer is $4\times 1$ and it should come with `double-stripe' AFM (dAFM) spin order. It is consistent with a group of previous experiments although the experimental identification of magnetic configuration remains being awaited. Based on this understanding we further explore the possible strain engineering of ground state properties. By applying tensile strain, monolayer VTe$_2$ exhibits two successive transitions from dAFM $4\times 1$ CDW to dAFM $2\times 1$, and then eventually to ferromagnetic (FM) $2\times 1$ phase. Our study not only successfully resolves the controversy regarding the ground state configuration of VTe$_2$ monolayer by unveiling the `hidden' role of underlying magnetism, but it also provides a useful way to control it and to achieve another monolayer ferromagnetism.

\section{Results and discussion}

\begin{figure*}[t]
	\centering
	\includegraphics[width=1.0\linewidth]{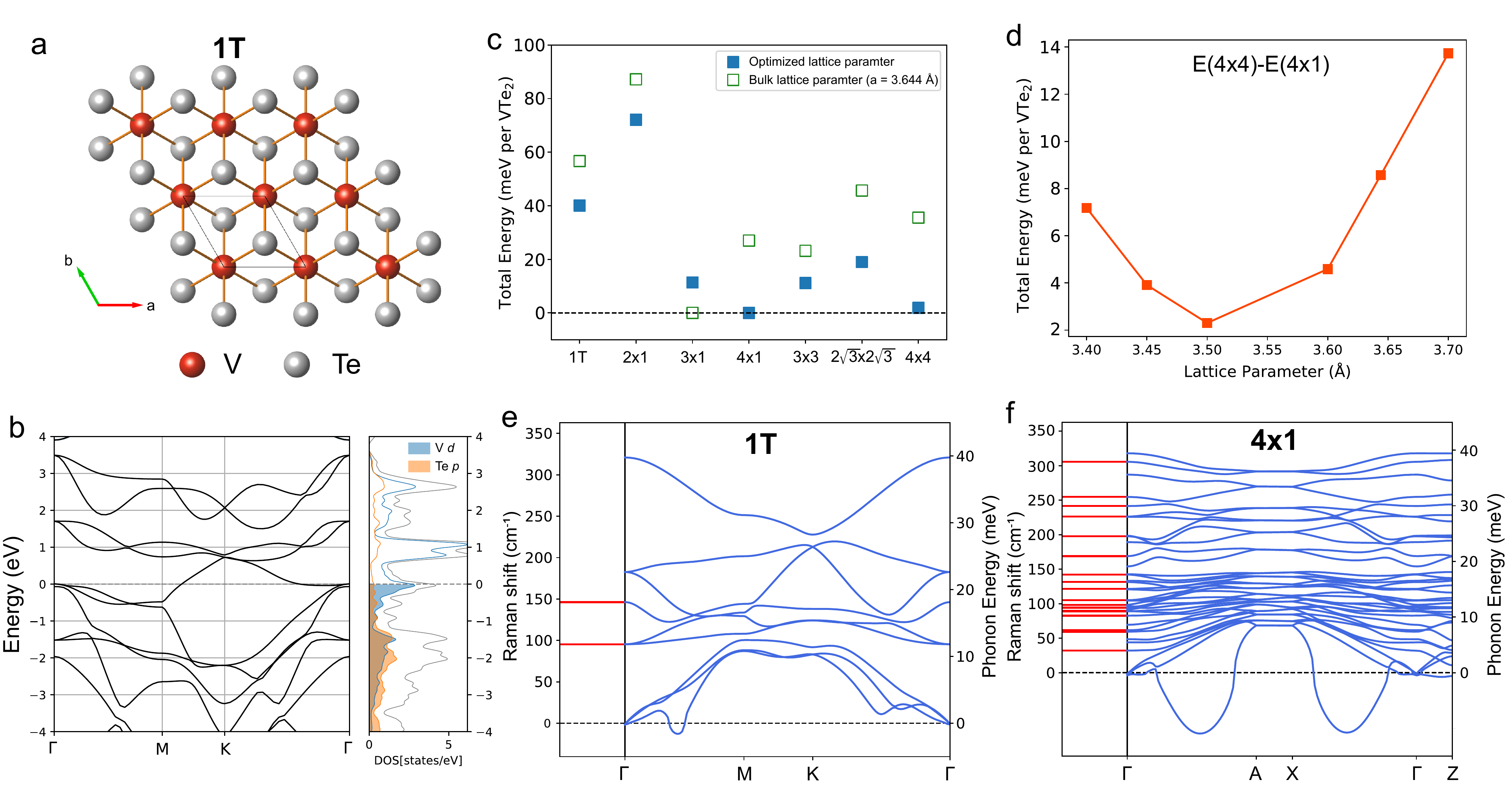}
	\caption{(a) The crystal structure of 1T (P-3m1) monolayer VTe$_2$. The red and white spheres refer to V and Te atoms, respectively. (b)  The calculated band dispersion and the projected density of states. The blue and orange color shows V-$d$ and Te-$p$ projected states. The gray line shows the total density of states. The horizontal dashed line is $E_F$. (c)  The calculated total energies relative to the most stable phase. The filled blue and the empty green symbols are the results obtained, respectively, from the theoretically optimized and the bulk experimental lattice parameter  within which the atomic positions were optimized.  (d) The total energy difference between 4x4 and 4x1 CDW phase as a function of lattice parameter. (e \& f) The calculated phonon dispersions and Raman-active modes for (d) 1T and (e) 4x1 CDW monolayer.}
	\label{figure1}
\end{figure*}

Fig.~\ref{figure1}a shows the high symmetric 1T structure of VTe$_2$. Vanadium atoms are located in the Te octahedral cage and form a 2D triangular lattice. A recent single crystal study of bulk VTe$_2$ demonstrates that the $3\times1$ and $3\times3$ CDW phases can be stabilized in a controllable way and in collaboration with antiferromagnetism \cite{won2020polymorphic}. On the other hand, the situation of a 1T-based monolayer is quite controversial as briefly outlined above. Therefore, the correct ground state and the mechanism that stabilizes it are important open questions.

From the point of view of electronic structure, a noticeable difference between bulk and monolayer is the well developed van Hove singularity (vHs) at the Fermi level ($E_F$) as presented in Fig.~\ref{figure1}b. While  vHs is also found in monolayer VSe$_2$ \cite{kim2020VSe2}, the $\Gamma$ hole pocket is not observed in the case of VTe$_2$ monolayer \cite{sugawara_monolayer_2019,wang_evidence_2019,wong_metallic_2019, coelho_monolayer_2019}. There have been some discussions of the incommensurate nesting vector and its role in stabilizing  $4\times4$ CDW  \cite{sugawara_monolayer_2019,wang_evidence_2019}.

\subsection{Nonmagnetic monolayers}

To understand the ground state property of VTe$_2$ monolayer and to elucidate the origin of multiple CDW phases, we performed the total energy calculations. As a first step, we assumed the nonmagnetic phase because there has been no experimental report yet on any magnetic order in this material. Fig.~\ref{figure1}c presents a summary of our results. Here we considered two different lattice parameter choices; namely, the bulk experimental (open green symbols) and the theoretically optimized value (filled blue) while the internal atomic positions are fully relaxed in both cases. Monolayer VTe$_2$ is typically grown on a substrate of  graphene/SiC or highly oriented pyrolytic graphite (HOPG) whose lattice parameters of 3.6--3.7 \AA \cite{wang_evidence_2019, miao_real-space_2020,wu_orbital-collaborative_2020,wong_metallic_2019} are therefore not much different from bulk VTe$_2$ (3.644 \AA) \cite{won2020polymorphic}. 

At the experimental lattice parameter, the bulk-like \cite{bronsema1984crystal,won2020polymorphic} $3\times 1$ structure is found to be most stable and the $3\times 3$ is the next having the higher energy by $\sim$25 meV/f.u.. Note that 1T and $2\times 1$ phases are energetically unstable having much higher energy. Our result is therefore in good agreement with  previous theoretical studies of Ref.~\citenum{wong_metallic_2019, coelho_monolayer_2019}, but not with the experimental reports \cite{sugawara_monolayer_2019, wang_evidence_2019, wong_metallic_2019, coelho_monolayer_2019, miao_real-space_2020, liu_multimorphism_2020, wu_orbital-collaborative_2020}.

On the other hand, the result from the optimized lattice parameter gives rise to the 4-formula-unit CDW ground state as reported in some experiments \cite{wang_evidence_2019, wong_metallic_2019, coelho_monolayer_2019, miao_real-space_2020, liu_multimorphism_2020, wu_orbital-collaborative_2020}. As shown in Fig.~\ref{figure1}c, the $4\times 1$ and $4\times 4$ phases have the lowest energy, and their energy difference is less than 1.9 meV/f.u.. The $2\sqrt{3}\times2\sqrt{3}$ phase reported by Ref.~\citenum{liu_multimorphism_2020} has quite high energy \cite{liu_multimorphism_2020}. It is also important to note that the $2\times1$ CDW, which was reported in the bilayer and a few layers of VTe$_2$ \cite{ma_evidence_2012,dai_multiple_2019, coelho_monolayer_2019}, is not quite favorable energetically for the monolayer.

Our result demonstrates that good care must be taken in the simulation of  experimental situations. It is well known that conventional approximations adopted in DFT typically produce  systematic deviations in reproducing the experimental lattice property. For example, GGA-type approximation often overestimates the lattice parameter while LDA underestimates. It can therefore be important to note that adopting the experimental value can correspond to the situation that effectively imposes the contraction or enlargement of lattice rather than the optimized value for the given material. For such sensitive cases as VTe$_2$ and other transition-metal chalcogenides, in which multiple CDW phases and other competing orders are intertwined, this kind of computation details should be taken into account more seriously.

\begin{figure*}[t]
	\centering
	\includegraphics[width=0.9\linewidth]{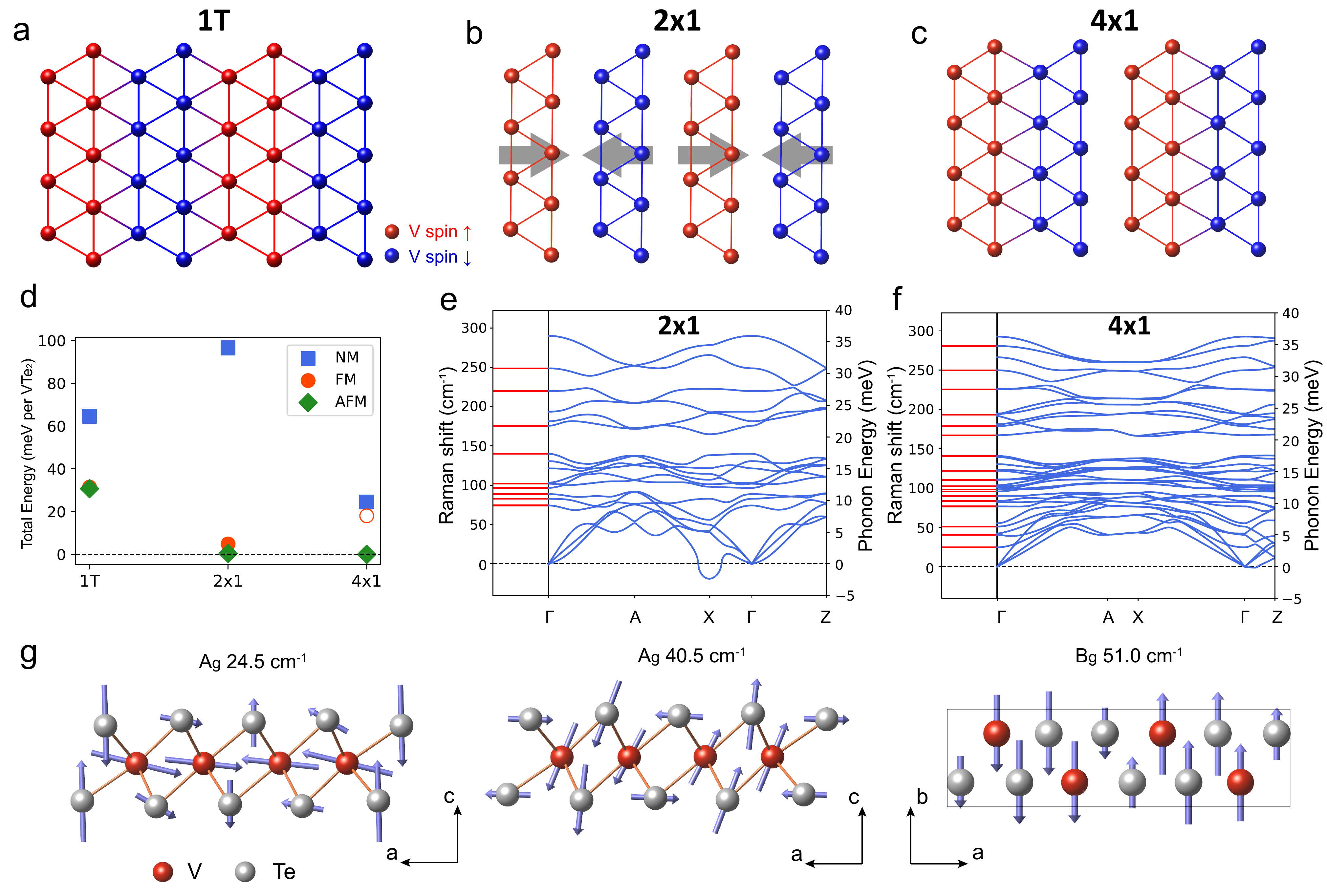}
	\caption{(a)--(c) The magnetic ground state for (a) 1T, (b) $2\times1$, and (c) $4\times1$ CDW phase. The red and blue spheres represent the up and down spin V sites, respectively. The gray arrows in (b) show the imaginary phonon mode at the \textrm{X} point depicted in (e). (d) The calculated total energies of nonmagnetic (blue squares), FM (red circles), and AFM (green diamonds) phase with respect to the most stable one. The open red circle for 4x1 structure indicates that the converged solution is not FM but ferrimagnetic carrying the net magnetic moment of 0.29~$\mu_B$ per unit cell. (e \& f) The calculated phonon dispersion (blue lines; right panel) and Raman-active modes (red lines; left panel) for the most stable (e) AFM $2\times1$ and (f) AFM $4\times1$ case. (g) The visualization of the three lowest energy Raman active modes for AFM $4\times1$ CDW. The arrows show the direction of atomic displacements.}
	\label{figure2}
\end{figure*}

We now examine the relative stabilization of $4\times1$ and $4\times4$ CDWs in detail. Fig.~\ref{figure1}d presents the calculated total energy difference as a function of the lattice parameter. The $4\times1$ phase is always more stable in the entire range. Interestingly, both structures have their minimum energies at $a=3.50$ \AA~ and the difference becomes minimized at that point. The smallness of this energy difference, corresponding to $\sim$26.7 K/f.u., likely indicates the possible control of the ground state by strain for example.

Important is to check the dynamical stability of each phase. The calculation results of  phonon dispersion are presented in Fig.~\ref{figure1}e and f. Note that imaginary phonon mode is clearly identified not only for 1T (Fig.~\ref{figure1}e) but also for the most stable $4\times1$ phase (Fig.~\ref{figure1}f). While the unstable 1T phonon observed in the middle of $\Gamma$--$\textrm{M}([1/2,0,0])$ line indicates the `$\times$4' modulations such as $4\times 1$ and $4\times 4$  \cite{wang_evidence_2019,liu_multimorphism_2020}, the imaginary  mode found in $4\times1$ (at the middle of $\textrm{X}[1/2,0,0]$--$\Gamma$) indicates the $4\times4$ modulations \cite{coelho_monolayer_2019}. Given that the energetically most stable $4\times1$ CDW phase is dynamically  unstable, the result implies that the ground state is not well described within the nonmagnetic picture, and requires further investigation.

\subsection{Magnetism}

Recently, the intriguing interplay of charge, spin, and lattice is found to play a key role in determining the polymorphic bulk ground state of VTe$_2$ \cite{won2020polymorphic}. For monolayers, on the other hand, the role of magnetism has not been explored. Considering that the direct  detection of magnetic order is experimentally  challenging for ultra-thin systems, particularly the AFM order, the intriguing possibility of magnetism and its `hidden' role interplaying with other degrees of freedom are certainly worthy of being investigated.

Here we examine the 1T, $2\times1$, and $4\times1$ structure together with FM and AFM spin order. 
We take both single-stripe AFM (sAFM) and dAFM order into account. The former is found to be a well converged solution only for 1T, and dAFM order is the ground state for 1T, $2\times1$, and $4\times1$; see Fig.~\ref{figure2}a--c.
The total energy results are summarized in Fig.~\ref{figure2}d. First of all, let us take note that magnetic orders significantly reduce the energy of the system. In 1T phase, FM and AFM orders have almost the same energy whereas the AFM order is clearly more stable in $2\times1$ and $4\times1$ CDW phases. Interestingly, the $2\times1$ phase is less favorable energetically than 1T unless magnetism is considered. Combined with AFM order, however, the $2\times1$ CDW phase becomes stabler than 1T by about 30 meV/f.u.. In the case of $4\times1$, we found, the initial FM order converges into a `ferrimagnetic' phase having the `up-down-down-up (↑-↓-↓-↑)'-type  order carrying a non-zero net moment. The most stable configuration is dAFM also for $4\times1$ whose energy is almost the same with that of dAFM $2\times1$ CDW ($\Delta E_{2\times1 - 4\times1}^{\rm dAFM}$ < 0.5 meV/VTe$_2$) which can be regarded as being consistent with previous experiment \cite{coelho_monolayer_2019, wong_metallic_2019, liu_multimorphism_2020}.

\begin{figure*}[t]
	\centering
	\includegraphics[width=1\linewidth]{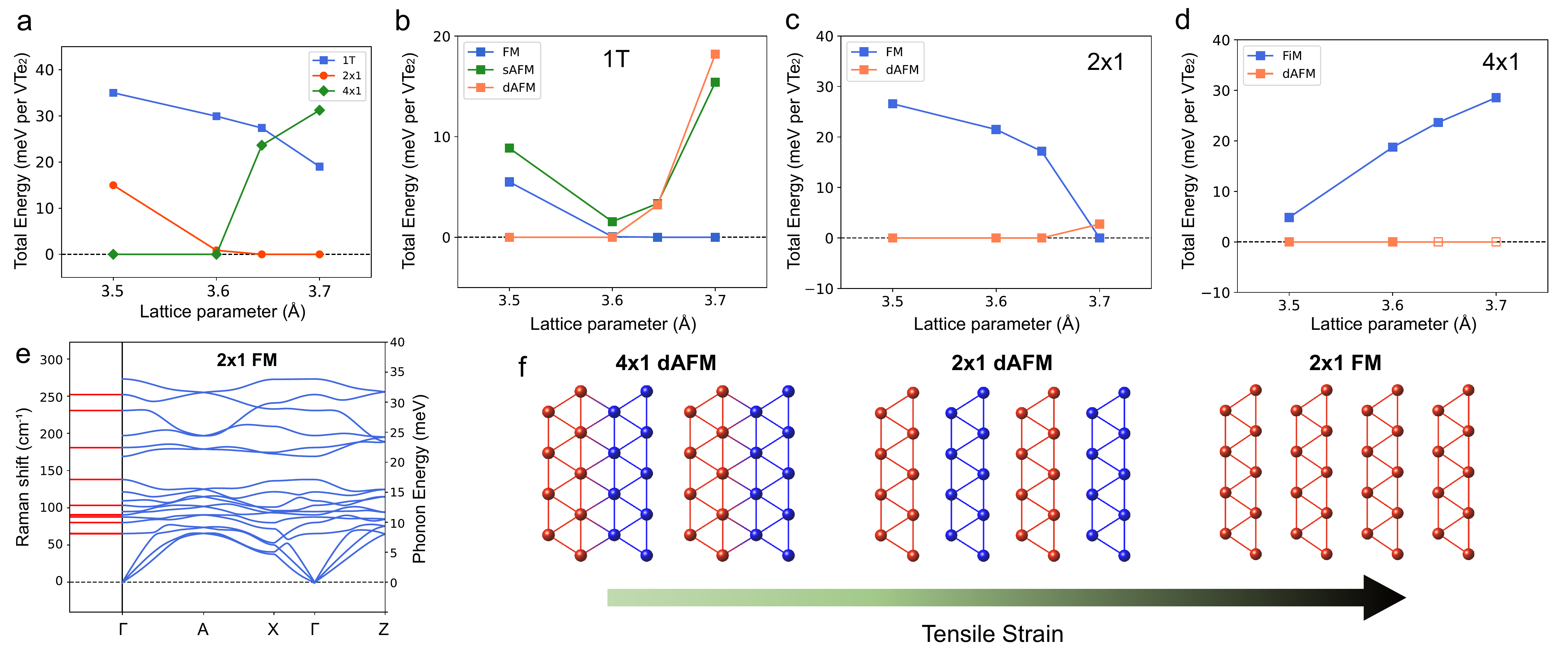}
	\caption{(a) The calculated total energy of three different CDW/structure  types as a function lattice parameter. Blue squares, red circles, and green diamonds represent 1T, $2\times 1$, and $4\times 1$, respectively. Each point represents the most stable spin configuration at a given lattice value. (b) Calculated relative total energy for FM and AFM states for 1T VTe$_2$ monolayer for different lattice parameters. (c) The calculated total energy of $2\times 1$ CDW phase as a function of the lattice parameter. The results of the two most stable spin configurations, namely FM and dAFM, are presented. (d) The calculated total energy of $4\times 1$ CDW phase as a function of lattice parameter. The results of the two most stable spin configurations are presented.  The red square symbols refer to the dAFM spin order (↑-↑-↓-↓). The blue squares show the result of another type of AFM order (↑-↓-↓-↑). Although it converges to a `ferrimagnetic' order, we found, the net moment is negligible (<0.1$\mu_B$/f.u.).
		The open squares at $a\geq$3.644 \AA~ indicate that, in this strain regime, the $4\times 1$ CDW phase converges into $2\times 1$ through the relaxation calculations while maintaining the dAFM spin order. (e) The calculated phonon dispersion and Raman-active modes for the predicted FM 2x1 CDW phase $a\geq$3.7 \AA.  (f) A schematic summary of successive phase transitions as strain varies. The left, middle and right figure refers to the representative region of $a=3.5$ (free-standing or zero strain), 3.644 (bulk experimental value) and 3.7 (tensile strain) \AA, respectively. }
	\label{figure3}
\end{figure*}

A remarkable difference between $2\times1$ and $4\times1$ phase is in their dynamic stability. Fig.~\ref{figure2}e and f show the calculated phonon dispersion. In $2\times1$ CDW phase, clearly noted is the imaginary phonon mode centered at X point ([1/2,0,0]), which indicates the structural modulation toward $4\times1$ type as schematically represented by arrows in Fig.~\ref{figure2}b. In fact, the phonon modes of $4\times1$ phase are all quite stable as shown in Fig.~\ref{figure2}f. It is important to note that  the dynamical stability is achieved only when we take the spin order into account (compare Fig.~\ref{figure2}f to Fig.~\ref{figure1}f), demonstrating the key role of magnetism in stabilizing CDW ground state. Our result thus provides an intriguing new example of `hidden' interplay of magnetism with charge and lattice in 2D materials.

As the conventional experimental techniques of determining spin order (e.g., neutron scattering) can be severely limited for 2D materials, Raman spectroscopy has been a useful alternative, albeit indirect, tool  \cite{kim2019suppression, lee2016ising, cenker2021direct}. To facilitate the possible future experimental study, we calculated Raman-active modes shown in Fig.~\ref{figure2}f; see the red lines in the left panel which can be directly compared with experiments. Due to the structural similarity, the  $4\times1$ modes are not much different from those of $2\times1$ phase (Fig.~\ref{figure2}e), and many states are located near 100 cm$^{-1}$. However, the states near $\leq$50 cm$^{-1}$  are found only  in $4\times1$ CDW phase. In fact, these modes are the vibrations that break the $2\times1$ structural symmetry and lead to the $4\times 1$ CDW. As depicted in Fig.~\ref{figure2}g, these three have the opposite direction atomic movements from the $2\times1$ chain. Thus, their experimental verification can be the key signature of $4\times1$ phase stabilized by `hidden' dAFM spin order.

\subsection{Controlling the ground state}

The active interplay of spin, charge, and lattice immediately indicates the exciting possibility of controlling material properties. As a concrete example, here we examined the strain effect. Fig.~\ref{figure3}a shows the calculated total energies with respect to the most stable magnetic and CDW configuration at a given lattice parameter. The blue square, red circle, and green diamond symbols represent the lowest energy spin order of 1T, $2 \times 1$, and $4 \times 1$ phase, respectively. It is clearly noted that the most stable CDW pattern changes from $4 \times 1$ to $2 \times 1$, and this phase transition occurs in between $a=$3.6 and 3.644 \AA.

Moreover, this charge-lattice pattern transition is followed by a magnetic transition as the lattice parameter further increases, and the system becomes eventually FM at $a\approx$ 3.7~\AA.  Fig.~\ref{figure3}c and d shows the relative stabilization energy of the two most stable spin orders within the $2 \times 1$ and $4 \times 1$ CDW phase, respectively.  While dAFM remains most stable for $4 \times 1$ in the entire range of strain values, the ground state  for  $2 \times 1$ changes from dAFM to FM. The calculated total energies for 1T structure are presented in Fig.~\ref{figure3}b.

Fig.~\ref{figure3}f summarizes the evolution of the ground state properties as a function of strain. The first transition is in between two CDW phases of $4 \times 1$ and $2 \times 1$ while both maintain the dAFM spin order. The second is a spin reordering transition from dAFM to FM within the $2 \times 1$ CDW.

Fig.~\ref{figure3}e presents the phonon dispersion of the predicted ground state of FM $2 \times 1$ phase under tensile strain. It confirms that this FM $2 \times 1$ phase is  dynamically stable.  We emphasize that this interesting possibility of controlling the ground state becomes feasible through the spin-lattice-charge coupling. Also, the `hidden' spin order (in the sense that an AFM order is hardly identified in the monolayer via conventional methods) would unveil itself once the system becomes FM (which is much easier by means of  magnetic circular dichroism (XMCD) and/or magneto-optical Kerr effect (MOKE) for example). To facilitate the experimental verification, we once again calculated the Raman spectra shown in the left panel of Fig.~\ref{figure3}e. It can serve as a useful guideline or prediction for future experiments.

\begin{figure*}[t]
	\centering
	\includegraphics[width=1.0\linewidth]{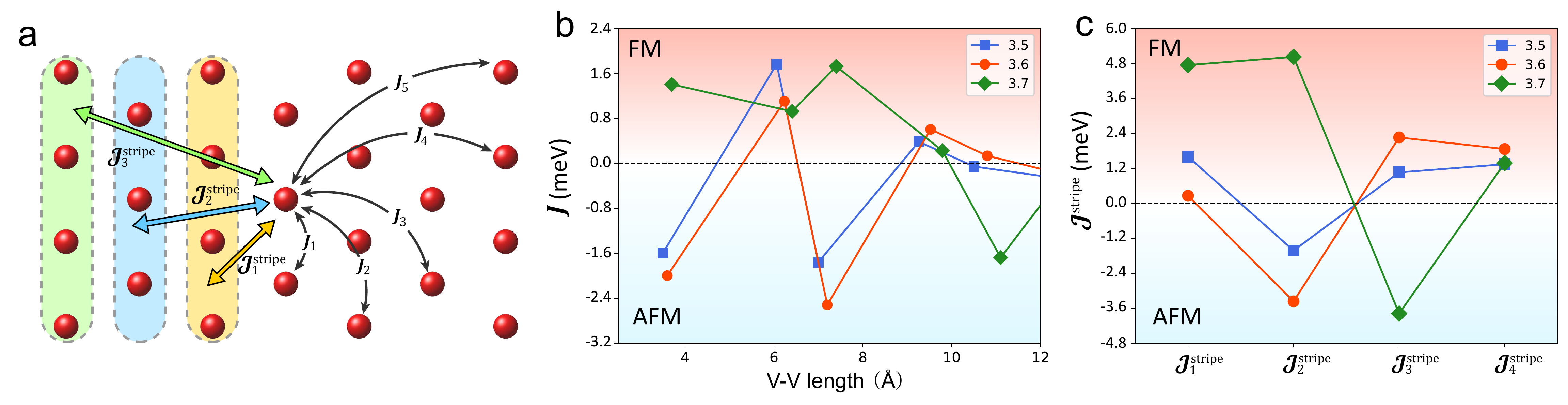}
	\caption{The strain effects for magnetic interactions on 1T VTe$_2$ monolayer. (a) The schematic figure representing magnetic interaction $J$ of $n$-th nearest neighbors and the net stripy interaction $\mathcal{J}^\text{stripe}$. (b \& c) The calculated (b) $J$ and (c) $\mathcal{J}^\text{stripe}$ values as a function of neighboring distances for different strain environments.}
	\label{figure4}
\end{figure*}

\subsection{Magnetic interactions}

As presented in Fig.~\ref{figure2}, the spin structure of 1T, $2\times1$, and $4\times1$ are basically similar although the CDW makes the inter-atomic or inter-`stripe' spacing different. To understand the magnetic interactions in greater detail, we performed the magnetic force linear response calculation \cite{Yoon_Reliability2018,oguchi1983magnetism,Antropov_spindynamics1996PRB,liechtenstein1987local,Katsnelson_magnetic_2000_PRB,Han_Electronic_PRB2004,YOON2020106927}. Here we fix the lattice symmetry to the most symmetric 1T because it provides the most useful insight for the underlying origin of the spin ground state evolution. The convention of our spin Hamiltonian is $\mathcal{H} = -\sum J_{i,j} \bf{e_i}\cdot \bf{e_j}$ where $\bf{e_{i,j}}$ refers to the normalized unit spin vector and $i, j$ represents the atomic sites. The results are summarized in Fig.\ref{figure4}b. At the lattice parameter $a=3.5$ (blue squares) and $3.6$ \AA~ (red circles), the first and the third neighboring $J$'s are AFM  (negative sign) while the second is FM (positive sign). Note that this exchange interaction profile makes the dAFM order stable. As $a$ further increases to 3.7 \AA~ (green diamonds), $J_{1,2,3}$ become all  FM. This result is therefore in good agreement with the predicted ground state evolution shown in Fig.~\ref{figure3}f. Our results show that the sign changes of $J_{1}$ and  $J_{3}$ are the key for the AFM to FM transition induced by tensile strain.

We also estimate the `inter-stripe' coupling $\mathcal{J}_n^{\text{stripe}}$ which represents the magnetic interaction of a given V site with its $n$-th neighboring `stripe' as depicted in Fig.~\ref{figure4}a. Note that a single index $n$ is enough because $\mathcal{J}_n^{\text{stripe}}$ should be the same for any choice of V site due to the high symmetric 1T structure. From the lattice structure, this atom-`stripe' interaction can be expressed as follow:
\begin{align}
	\mathcal{J}_1^{\text{stripe}}& = 2J_1 + 2J_2 + 2J_4 +\dots\\
	\mathcal{J}_2^{\text{stripe}}& = J_2 + 2J_3 + 2J_4 +\dots\\
	\mathcal{J}_3^{\text{stripe}}& = 2J_4 + 2J_5 +\dots.
\end{align}
The results are summarized in Fig.~\ref{figure4}c. At $a\leqq3.6$ \AA, $\mathcal{J}_1^\text{stripe}$ has a small positive value, indicating the FM coupling while $\mathcal{J}_2^\text{stripe}$ is AFM. As the lattice parameter increases up to $a=3.7$ \AA, $\mathcal{J}_1^\text{stripe}$ and $\mathcal{J}_2^\text{stripe}$ change to become positive, stabilizing the FM ground state, which is again in support of our prediction of magnetic phase transition. The rough estimation of FM critical temperature simply by summing up all couplings is  $T_{\textrm{c}} \simeq  \sum J_{ij} \bf{e_i}\cdot \bf{e_j} \simeq \textrm{216}$ K.

\section{Conclusions}

The comprehensive first-principles calculations are carried out to resolve the controversy regarding the ground state property of monolayer VTe$_2$. Unveiled is the underlying key role of magnetism which has been ‘hidden’ from  experiments. We found that the $4\times1$ CDW is the most stable charge-lattice configuration, which is the first consistent theoretical result with the experiments. Further, we show that this phase can be stable only when it comes with dAFM spin order. Finally, based on this intriguing interplay of spin, charge, and lattice, we suggest possible strain engineering. By applying tensile strain, one can induce two successive transitions, first from dAFM $4\times1$ to $2\times1$ CDW phase, and then eventually to FM order. It renders VTe$_2$ an exciting candidate for monolayer ferromagnet. Our results shed new light on understanding the magnetism in 2D materials and provide a useful strategy to control their properties.

\section{Methods}

DFT calculations were carried out using the Vienna \textit{ab initio} simulation package (VASP) \cite{Kresse_PhysRevB1996} based on the projector augmented wave (PAW) potential \cite{Blochl_PhysRevB1994} and within the Perdew-Burke-Ernzerhof (PBE) type of generalized gradient approximation (GGA) functional \cite{Perdew_Burke_Ernzerhof}. The wave functions were expanded with plane waves up to an energy cutoff of 400 eV, and gamma-centered k-meshes of more than 70 points per \AA~ were adopted except for the $c$ direction. The vacuum space is more than 20 \AA. The lattice parameters and the internal coordinates were fully optimized with the force criterion of 10$^{-3}$ eV/\AA. The strain effect was taken into account by optimizing the internal coordinates with the given fixed lattice constants. The PHONOPY \cite{TOGO20151} package was used to calculate phonon and Raman spectra. In phonon calculations, $4\times4\times1$ and $1\times4\times1$ supercell was used for the nonmagnetic 1T (P-3m1) and  the magnetic structures, respectively. Magnetic force linear response calculations \cite{Yoon_Reliability2018,oguchi1983magnetism,Antropov_spindynamics1996PRB,liechtenstein1987local,Katsnelson_magnetic_2000_PRB,Han_Electronic_PRB2004,YOON2020106927} were performed to estimate the interatomic magnetic exchange interactions. The interatomic Hamiltonian is obtained with Crystal Orbital Hamilton Populations (COHP) within LOBSTER code \cite{nelson2020lobster}. In this computation, we used the FM electronic structure  as our main data for convenience after the double-check that the results from AFM ones show the same tendency.

\section*{Notes}
The authors declare no competing financial interest.

\begin{acknowledgement}

This work was supported by the National Research Foundation of Korea (NRF) grant funded by the Korea government (MSIT) (No. 2021R1A2C1009303 and No. NRF2018M3D1A1058754). This research was supported by the KAIST Grand Challenge 30 Project (KC30) in 2021 funded by the Ministry of Science and ICT of Korea and KAIST (N11210105).

\end{acknowledgement}

\providecommand{\latin}[1]{#1}
\makeatletter
\providecommand{\doi}
{\begingroup\let\do\@makeother\dospecials
	\catcode`\{=1 \catcode`\}=2 \doi@aux}
\providecommand{\doi@aux}[1]{\endgroup\texttt{#1}}
\makeatother
\providecommand*\mcitethebibliography{\thebibliography}
\csname @ifundefined\endcsname{endmcitethebibliography}
{\let\endmcitethebibliography\endthebibliography}{}

\end{document}